\documentclass[final,5p,times,twocolumn,longtitle]{elsarticle}
\usepackage{amssymb}
\usepackage{amsmath}
\usepackage{url}
\usepackage{doi}
\usepackage{graphicx}
\usepackage{lineno}
\usepackage{orcidlink}
\usepackage{hyperref}
\journal{Nuclear Instruments and Methods in Physics Research - section A (NIM-A)}
\date{November 2024}

\begin{document}
\begin{frontmatter}
\title{CMS RPC Non-Physics Event Data Automation Ideology}
\author[6]{A.~Dimitrov~\orcidlink{0000-0003-2899-701X}}
\ead{anton.dimitrov@cern.ch}

\affiliation[1]{organization={Vrije Universiteit Brussel}, city={Brussel}, country={Belgium}}
\author[1]{M.~Tytgat~\orcidlink{0000-0002-3990-2074}\fnref{tytgat}}
\fntext[tytgat]{Also at Ghent University, Ghent, Belgium}

\affiliation[2]{organization={Universiteit Gent}, city={Gent}, country={Belgium}}
\author[2]{K.~Mota~Amarilo\orcidlink{0000-0003-1707-3348}\fnref{amarilo}}
\fntext[amarilo]{Now at UERJ, Rio de Janeiro, Brazil}
\author[2]{A.~Samalan~\orcidlink{0000-0001-9024-2609}\fnref{samalan}}
\fntext[samalan]{Now at PSI, Villigen, Switzerland} 
\author[2]{K.~Skovpen~\orcidlink{0000-0002-1160-0621}}

\affiliation[3]{organization={Centro Brasileiro de Pesquisas Fisicas}, city={Rio de Janeiro}, country={Brazil}}
\author[3]{G.A.~Alves~\orcidlink{0000-0002-8369-1446}}
\author[3]{E.~Alves~Coelho~\orcidlink{0000-0001-6114-9907}}
\author[3]{F.~Marujo da Silva~\orcidlink{0000-0002-3555-0489}}

\affiliation[4]{organization={Universidade do Estado do Rio de Janeiro}, city={Rio de Janeiro}, country={Brazil}}
\author[4]{M.~Barroso~Ferreira~Filho~\orcidlink{0000-0003-3904-0571}}
\author[4]{E.M.~Da~Costa~\orcidlink{0000-0002-5016-6434}}
\author[4]{D.~De~Jesus~Damiao~\orcidlink{0000-0002-3769-1680}}
\author[4]{S.~Fonseca~De~Souza~\orcidlink{0000-0001-7830-0837}}
\author[4]{R.~Gomes~De~Souza~\orcidlink{0000-0003-4153-1126}}
\author[4]{L.~Mundim~\orcidlink{0000-0001-9964-7805}}
\author[4]{H.~Nogima~\orcidlink{0000-0001-7705-1066}}
\author[4]{J.P.~Pinheiro~\orcidlink{0009-0003-6293-3332}}
\author[4]{A.~Santoro~\orcidlink{0000-0002-0568-665X}}
\author[4]{M.~Thiel~\orcidlink{0000-0001-7139-7963}}

\affiliation[5]{organization={Institute for Nuclear Research and Nuclear Energy, Bulgarian Academy of Sciences}, city={Sofia}, country={Bulgaria}}
\author[5]{A.~Aleksandrov~\orcidlink{0000-0001-6934-2541}}
\author[5]{R.~Hadjiiska~\orcidlink{0000-0003-1824-1737}}
\author[5]{P.~Iaydjiev~\orcidlink{0000-0001-6330-0607}}
\author[5]{M.~Shopova~\orcidlink{0000-0001-6664-2493}}
\author[5]{G.~Sultanov~\orcidlink{0000-0002-8030-3866}}

\affiliation[6]{organization={Faculty of Physics, University of Sofia}, city={Sofia}, country={Bulgaria}}
\author[6]{L.~Litov~\orcidlink{0000-0002-8511-6883}}
\author[6]{B.~Pavlov~\orcidlink{0000-0003-3635-0646}} 
\author[6]{P.~Petkov~\orcidlink{0000-0002-0420-9480}} 
\author[6]{A.~Petrov~\orcidlink{0009-0003-8899-1514}} 
\author[6]{E.~Shumka~\orcidlink{0000-0002-0104-2574}}

\affiliation[7]{organization={Institute of High Energy Physics and University of the Chinese Academy of Sciences}, city={Beijing}, country={China}}
\author[7]{P.~Cao}
\author[7]{W.~Diao}
\author[7]{Q.~Hou}
\author[7]{H.~Kou~\orcidlink{0000-0003-4927-243X}}
\author[7]{Z.-A.~Liu~\orcidlink{0000-0002-2896-1386}} 
\author[7]{J.~Song}
\author[7]{J.~Zhao~\orcidlink{0000-0001-8365-7726}}

\affiliation[8]{organization={School of Physics, Peking University}, city={Beijing}, country={China}}
\author[8]{S.J.~Qian~\orcidlink{0000-0002-0630-481X}}

\affiliation[9]{organization={Universidad de Los Andes}, city={Bogota}, country={Colombia}} 
\author[9]{C.~Avila~\orcidlink{0000-0002-5610-2693}}
\author[9]{D.A.~Barbosa~Trujillo~\orcidlink{0000-0001-6607-4238}} 
\author[9]{A.~Cabrera~\orcidlink{0000-0002-0486-6296}}
\author[9]{C.A.~Florez~\orcidlink{0000-0002-3222-0249}}
\author[9]{J.A.~Reyes~Vega}%

\affiliation[10]{organization={Physics Department, Faculty of science, Helwan University}, city={Cairo}, country={Egypt}}
\affiliation[12]{organization={The British University in Egypt}, city={Cairo}, country={Egypt}}
\author[10,12]{R.Aly~\orcidlink{0000-0001-6808-1335}\fnref{aly}}
\fntext[aly]{Also at Academy of Scientific Research and Technology of the Arab Republic of Egypt, Egyptian Network of High Energy Physics, Cairo, Egypt}

\affiliation[11]{organization={Department of Physics, Faculty of Science, Ain Shams University}, city={Cairo}, country={Egypt}}
\author[11]{A.~Radi~\orcidlink{0000-0002-7857-3445}\fnref{radi}}
\fntext[radi]{Also at Sultan Qaboos University, Muscat, Oman}

\author[12]{Y.~Assran~\orcidlink{0000-0001-6974-9595}\fnref{assran}}
\fntext[assran]{Also at Suez University, Suez, Egypt}

\affiliation[13]{organization={Center for High Energy Physics (CHEP-FU), Fayoum University}, city={El-Fayoum}, country={Egypt}}
\author[13]{I.~Crotty}%
\author[13]{M.A.~Mahmoud~\orcidlink{0000-0001-8692-5458}}

\affiliation[14]{organization={Institut de Physique des 2 Infinis de Lyon}, city={Villeurbanne}, country={France}}
\author[14]{M.~Gouzevitch~\orcidlink{0000-0002-5524-880X}}
\author[14]{G.~Grenier~\orcidlink{0000-0002-1976-5877}}
\author[14]{I.B.~Laktineh~\orcidlink{0000-0003-1394-3158}} 
\author[14]{L.~Mirabito~\orcidlink{0009-0009-6209-1968}}

\affiliation[15]{organization={Georgian Technical University}, city={Tbilisi}, country={Georgia}}
\author[15]{I.~Bagaturia~\orcidlink{0000-0001-8646-4372}}
\author[15]{I.~Lomidze~\orcidlink{0009-0002-3901-2765}}
\author[15]{Z.~Tsamalaidze~\orcidlink{0000-0001-5377-3558}\fnref{tsamalaidze}}
\fntext[tsamalaidze]{Also at an institute or an international laboratory covered by a cooperation agreement with CERN} 

\affiliation[16]{organization={Institute for Research in Fundamental Sciences}, city={Tehran}, country={Iran}}
\author[16]{V.~Amoozegar}  
\author[16]{B.~Boghrati~\orcidlink{0009-0006-4923-6315}} 
\author[16]{M.~Ebrahimi~\orcidlink{0000-0002-8967-7725}} 
\author[16]{F.~Esfandi~\orcidlink{0009-0007-9245-0840}} 
\author[16]{Y.~Hosseini~\orcidlink{0000-0001-8179-8963}}
\author[16]{M.~Mohammadi Najafabadi~\orcidlink{0000-0001-6131-5987}} 
\author[16]{E.~Zareian~\orcidlink{0009-0003-9123-344X}}

\affiliation[17]{organization={INFN Sezione di Bari}, city={Bari}, country={Italy}}
\affiliation[171]{organization={Università  di Bari}, city={Bari}, country={Italy}}
\affiliation[172]{organization={Politecnico di Bari}, city={Bari}, country={Italy}}
\author[17,171]{M.~Abbrescia~\orcidlink{0000-0001-8727-7544}}
\author[17,172]{N.~De~Filippis~\orcidlink{0000-0002-0625-6811}}
\author[17,172]{G.~Iaselli~\orcidlink{0000-0003-2546-5341}}
\author[17]{F.~Loddo~\orcidlink{0000-0001-9517-6815}}
\author[17,172]{G.~Pugliese~\orcidlink{0000-0001-5460-2638}}
\author[17]{D.~Ramos~\orcidlink{0000-0002-7165-1017}}

\affiliation[18]{organization={INFN Laboratori Nazionali di Frascati}, city={Frascati}, country={Italy}}
\author[18]{L.~Benussi~\orcidlink{0000-0002-2363-8889}}
\author[18]{S.~Bianco~\orcidlink{0000-0002-8300-4124}}
\author[18]{S.~Meola~\orcidlink{0000-0002-8233-7277}\fnref{meola}}
\fntext[meola]{Also at Universit\`{a} degli Studi Guglielmo Marconi, Roma, Italy}
\author[18]{D.~Piccolo~\orcidlink{0000-0001-5404-543X}}

\affiliation[19]{organization={INFN Sezione di Napoli}, city={Napoli}, country={Italy}}
\affiliation[191]{organization={Università di Napoli 'Federico II'}, city={Napoli}, country={Italy}}
\affiliation[192]{organization={Dipartimento di Ingegneria Elettrica e delle Tecnologie dell'Informazione - Università Degli Studi di Napoli Federico II}, city={Napoli}, country={Italy}}
\author[19]{S.~Buontempo~\orcidlink{0000-0001-9526-556X}}
\author[19,191]{F.~Carnevali~\orcidlink{0000-0003-3857-1231}}
\author[192]{F.~Fienga~\orcidlink{0000-0001-5978-4952}}
\author[19,191]{L.~Lista~\orcidlink{0000-0001-6471-5492}\fnref{lista}}
\fntext[lista]{Also at Scuola Superiore Meridionale, Universit\`{a} di Napoli 'Federico II', Napoli, Italy}
\author[19]{P.~Paolucci~\orcidlink{0000-0002-8773-4781}\fnref{paolucci}}
\fntext[paolucci]{Also at CERN, European Organization for Nuclear Research, Geneva, Switzerland}

\affiliation[20]{organization={INFN Sezione di Pavia}, city={Pavia}, country={Italy}}
\affiliation[201]{organization={Università di Pavia}, city={Pavia}, country={Italy}}
\author[20]{A.~Braghieri~\orcidlink{0000-0002-9606-5604}}
\author[20,201]{P.~Montagna~\orcidlink{0000-0001-9647-9420}}
\author[20,201]{C.~Riccardi~\orcidlink{0000-0003-0165-3962}}
\author[20]{P.~Salvini~\orcidlink{0000-0001-9207-7256}}
\author[20,201]{P.~Vitulo~\orcidlink{0000-0001-9247-7778}}

\affiliation[21]{organization={Hanyang University}, city={Seoul}, country={Korea}}
\author[21]{E.~Asilar~\orcidlink{0000-0001-5680-599X}}
\author[21]{T.J.~Kim~\orcidlink{0000-0001-8336-2434}}
\author[21]{Y.~Ryou~\orcidlink{0009-0002-2762-8650}}

\affiliation[22]{organization={Korea University}, city={Seoul}, country={Korea}}
\author[22]{S.~Choi~\orcidlink{0000-0001-6225-9876}}
\author[22]{B.~Hong~\orcidlink{0000-0002-2259-9929}}
\author[22]{K.S.~Lee~\orcidlink{0000-0002-3680-7039}}

\affiliation[23]{organization={Kyung Hee University, Department of Physics}, city={Seoul}, country={Korea}}
\author[23]{J.~Goh~\orcidlink{0000-0002-1129-2083}}
\author[23]{J.~Shin~\orcidlink{0009-0004-3306-4518}}

\affiliation[24]{organization={Sungkyunkwan University}, city={Suwon}, country={Korea}}
\author[24]{Y.~Lee~\orcidlink{0000-0001-6954-9964}}

\affiliation[25]{organization={Benemerita Universidad Autonoma de Puebla}, city={Puebla}, country={Mexico}}
\author[25]{I.~Pedraza~\orcidlink{0000-0002-2669-4659}}
\author[25]{C.~Uribe~Estrada~\orcidlink{0000-0002-2425-7340}}

\affiliation[26]{organization={Centro de Investigacion y de Estudios Avanzados del IPN}, city={Mexico City}, country={Mexico}} 
\author[26]{H.~Castilla-Valdez~\orcidlink{0009-0005-9590-9958}}
\author[26]{R.~Lopez-Fernandez~\orcidlink{0000-0002-2389-4831}}
\author[26]{A.~S\'{a}nchez~Hern\'{a}ndez~\orcidlink{0000-0001-9548-0358}}

\affiliation[27]{organization={Universidad Iberoamericana}, city={Mexico City}, country={Mexico}}
\author[27]{M.~Ram\'{i}rez~Garc\'{i}a~\orcidlink{0000-0002-4564-3822}}
\author[27]{D.L.~Ramirez Guadarrama~\orcidlink{0000-0001-7280-8269}}
\author[27]{M.A.~Shah~\orcidlink{0009-0003-0581-9090}} 
\author[27]{E.~Vazquez~\orcidlink{0000-0001-6379-3982}}
\author[27]{N.~Zaganidis}%

\affiliation[28]{organization={National Centre for Physics, Quaid-I-Azam University}, city={Islamabad}, country={Pakistan}}
\author[28]{A.~Ahmad~\orcidlink{0000-0002-4770-1897}}
\author[28]{M.I.~Asghar~\orcidlink{0000-0002-7137-2106}} 
\author[28]{H.R.~Hoorani~\orcidlink{0000-0002-0088-5043}}
\author[28]{S.~Muhammad}%

\affiliation[29]{organization={Massachusetts Institute of Technology}, city={Cambridge, Massachusetts}, country={USA}}
\author[29]{J.~Eysermans~\orcidlink{0000-0001-6483-7123}}

\author[]{\\on behalf of the CMS Collaboration}

    \begin{abstract}
       This paper presents a streamlined framework for real-time processing and analysis of condition data from the CMS experiment Resistive Plate Chambers (RPC). Leveraging data streaming, it uncovers correlations between RPC performance metrics, like currents and rates, and LHC luminosity or environmental conditions. The Java-based framework automates data handling and predictive modeling, integrating extensive datasets into synchronized, query-optimized tables. By segmenting LHC operations and analyzing larger virtual detector objects, the automation enhances monitoring precision, accelerates visualization, and provides predictive insights, revolutionizing RPC performance evaluation and future behavior modeling.  
    \end{abstract}
\end{frontmatter}

\section{Introduction}
The Compact Muon Solenoid (CMS) \cite{cms} experiment at CERN is one of the largest and most complex particle detectors in operation, designed to explore the fundamental properties of matter and energy at the Large Hadron Collider (LHC). Among its key components is the muon system \cite{cms_muon_det,cms_muonsys}, responsible for identifying and measuring the momentum of muons, particles that serve as essential signatures for various valuable physics processes.

Within the muon system, Resistive Plate Chambers (RPC) \cite{cms_rpc_abbrescia} play a pivotal role, offering rapid, efficient, and highly reliable triggering capabilities. These detectors comprise two high-resistivity bakelite electrodes, held parallel by a spacer mesh to form a thin, physical gap. Signals are collected by a copper-stripped readout plane positioned between two such gaps, enabling double-gap (DG) operation. Filled with a finely tuned gas mixture of 95.2\% tetrafluoroethane, 4.5\% isobutane, and 0.3\% sulfur hexafluoride, the detectors achieve an optimal balance of efficiency, cluster size, timing resolution, and durability essential for their performance \cite{cms_rpc_pugliese}.

Advancements in automation and machine learning are revolutionizing condition data monitoring and RPC detector performance analysis, elevating it into a domain of unprecedented precision, speed, and adaptability.
\section{\label{sec:impl} Automation Implementation}
The CMS RPC condition data automation framework, initially developed in Python and now implemented in Java for its second version, comprises over 40 automata categorized into main and auxiliary types, each meticulously designed for specific tasks.

\begin{figure}[htbp!]
\includegraphics[width=0.99\columnwidth]{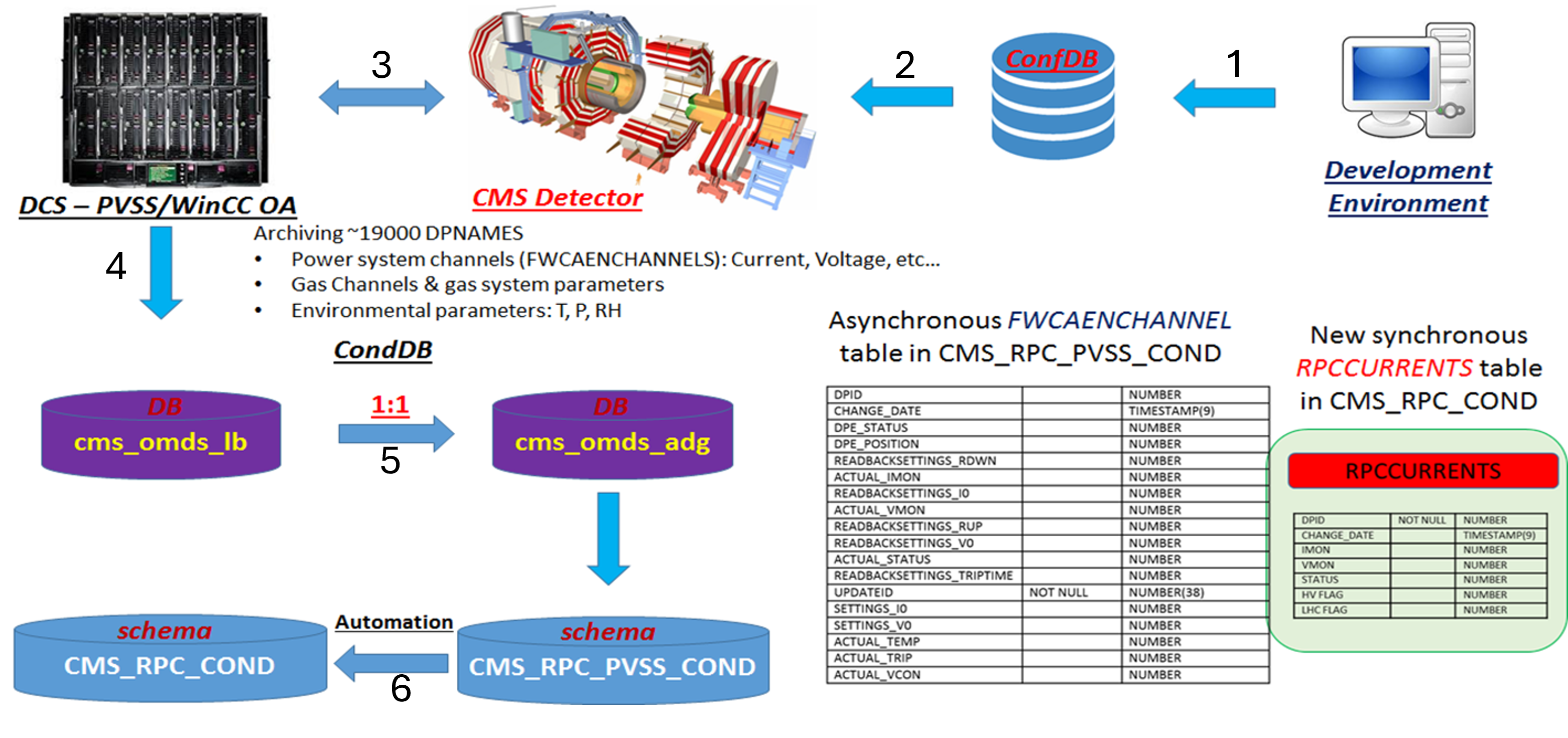}
\caption{ \label{fig:real} Overview of the RPC configuration and condition data flow: starting with the emulator (top right), which prepares and stores configuration parameters in ConfDB (1) that are loaded on the hardware via recipes during the CMS detector configuration (2). The RPC Detector Control System, DCS, monitors detector parameters and archives non-physics event data (3,4,5). The automation framework retrieves and processes this data, storing the results in structured formats for analysis and fast visualization (6). }
\end{figure}

This framework is deeply interwoven with the operation of the RPC detector and the Detector Control System (DCS) \cite{rpc_dcs}, which monitors and archives condition data within the Online Master Data Storage (\textit{OMDS}) instance of the \textit{CMSONR} production database \cite{cms_conddb}. \textit{CMSONR} refers to the computing framework responsible for managing the CMS Online Resources, which are essential for supporting the real-time operations of the CMS detector. The \textit{OMDS} database is an integral part of the data management infrastructure used in the CMS experiment. It is primarily designed to support the DCS and Online systems by storing and managing various operational data needed for the proper functioning and monitoring of the experiment. The entire configuration and data flow are illustrated in Fig. \ref{fig:real}. It begins with the emulator in the development environment, responsible for preparing and setting configuration parameters for the detector. These parameters are stored in the Configuration database as illustrated in Fig. \ref{fig:real} - data flow (1). All configuration parameters are loaded on the hardware during the CMS detector configuration - Fig. \ref{fig:real} - data flow (2). The RPC DCS plays a pivotal role, monitoring all detector parameters and archiving non-physics event data into the Condition database  \cite{cms_conddb, nga} - Fig. \ref{fig:real} - data flow (3,4).

The NextGen Archiver (NGA) \cite{nga} registers raw data outside a predefined bandwidth into the \textit{cms\_omds\_lb} database -  Fig. \ref{fig:real} - data flow (4). A real-time copy of this data is immediately transferred to the \textit{cms\_omds\_adg} read-only database \cite{cms_conddb} -  Fig. \ref{fig:real} - data flow (5). The RPC automation framework retrieves this raw data from the latter, processes it through a data streaming analysis pipeline, and stores the final results in the \textit{CMS\_RPC\_COND} schema within the production database -  Fig. \ref{fig:real} - data flow (6). This streamlined process ensures the accurate and efficient handling of detector condition data, supporting the operational integrity and research objectives of the CMS RPC system.

The main automata synchronize asynchronous data from the \textit{CMS\_RPC\_PVSS\_COND} schema into well-structured tables within the \textit{CMS\_RPC\_COND} schema on the production database. This includes raw data for RPC currents, LHC machine and beam modes, environmental conditions, RPC gas flows, and RPC rate data.

The auxiliary automata play a complementary role, segmenting the LHC filling cycle into four defined blocks to standardize condition data for analysis. They also facilitate in-depth investigations, including integrated charge accumulation, current evolution, high voltage (HV) conditioning, and dependencies of RPC currents and rates on LHC luminosity and charge per hit. These analyses extend to virtual objects representing lower-granularity detector components, such as regions, wheels, disks, stations, and sectors.

Operating on a 4-hour cycle, the framework has successfully processed vast datasets from the CMS experiment, enabling continuous monitoring and detailed analysis of RPC parameters. By integrating predictive modeling and virtual detector objects, the automation framework will further enhance our understanding of detector behavior, provide insights into performance trends, and support future operational forecasting.
\section{Project Rationale and Motivation}
The RPC condition data automation project began in 2016 to enhance studies of RPC detector performance. The initial motivation was to investigate the dependence of RPC currents on physical and environmental factors such as gas flow, temperature and relative humidity, and support aging studies, by integrating currents into charge, to ensure the long-term reliability of the detector. Over time, the project expanded to analyze current evolution and dependencies on LHC instantaneous luminosity. By 2023–2024, it shifted focus from hardware-specific channels, such as data point identifiers (DPID) \cite{rpc_dcs} of HV channels and various environmental sensors, to chamber-based identifiers (ChamberID), introducing new automata to populate chamber-centric database tables.
\section{Methods}
This section outlines the primary methods utilized or developed for the RPC automation framework, including data streaming, multithreading, data synchronization and tagging, the current probe, block averaging probe, the LHC block concept, and luminosity methods. These techniques optimize automation processes, improving speed and addressing the initial challenges encountered.
\subsection{Current Probe Method}
A key innovation is the development of the "Current Probe" method, a lightweight process with minimal CPU load that serves as the backbone for the entire automation framework.

The first major challenge was the immense volume of data stored in the current raw data table (\textit{FWCAENCHANNEL}), which made it impractical to work with direct SQL queries. To address this issue, the solution processes data in small, manageable chunks by breaking queries into narrow time windows, requesting data one day at a time. This approach significantly reduces the database workload—from causing server crashes to just 0.2\% CPU load—while enabling frequent, high-volume queries and leveraging the database design to handle high-demand operations efficiently.
\subsection{\label{subsec:synch} Data Synchronization}
\begin{figure}[htbp!]
\includegraphics[width=0.99\columnwidth]{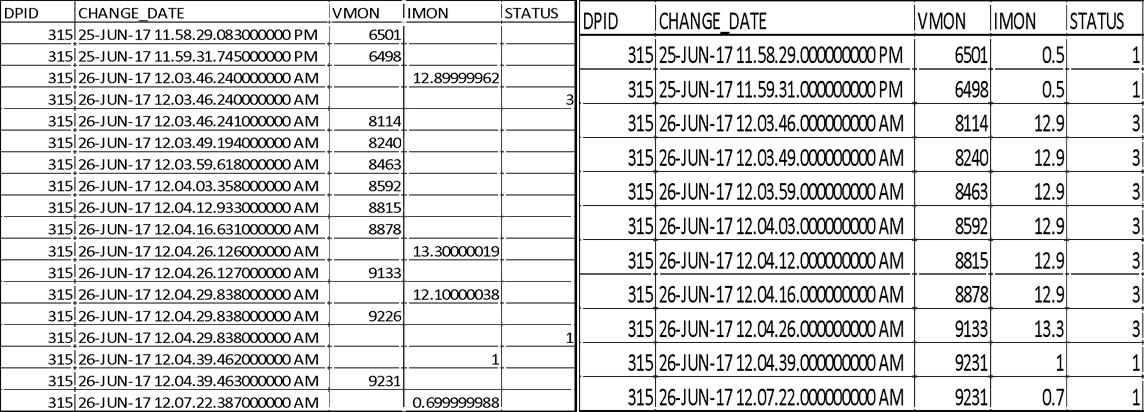}
\caption{ \label{fig:synch} A short, representative extract from the \textit{FWCAENCHANNEL} table in the \textit{CMS\_RPC\_PVSS\_COND} schema illustrates the raw data storage format, where only one column is populated per \textit{CHANGE\_DATE} timestamp (left). By aggregation within a second, the three fields are combined into a single row, and the "lead" analytical function fills in the voids until the next recorded entry per field is encountered. The synchronized data is stored in the \textit{RPCCURRENTS} table within the automation \textit{CMS\_RPC\_COND} schema (right). }
\end{figure}

The second major challenge stemmed from the asynchronous data archiving by the DCS archiving manager (RDB/NGA) \cite{nga}. Data is stored in a change-based format, logging only updated parameters—such as monitored current (Imon), monitored voltage (Vmon), or status—at each timestamp, while other fields remain void. However, analyzing current data requires all three parameters to be available for every timestamp. This issue is resolved by employing Oracle analytical and aggregation functions to accurately populate the void fields (see example in Fig. \ref{fig:synch}).
\subsection{\label{subsec:tagging} Data Tagging}
\begin{figure}[htbp!]
\includegraphics[width=0.99\columnwidth]{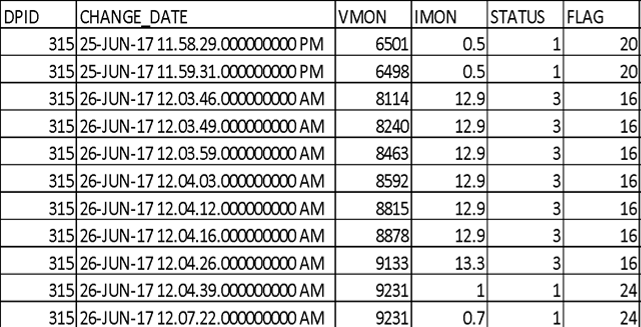}
\caption{ \label{fig:flag} An extended version of the \textit{RPCCURRENTS} table from Fig. \ref{fig:synch} is shown, with an additional column, FLAG, added to classify the current type. Using decimal representation of 6-bit encoding, the \textit{FLAG} column simplifies distinguishing between collision, CRAFT (cosmic currents at full magnetic field), CRUZET (cosmic currents at low magnetic field), standby, and offset currents.}
\end{figure}

\begin{figure}[htbp!]
\includegraphics[width=0.99\columnwidth]{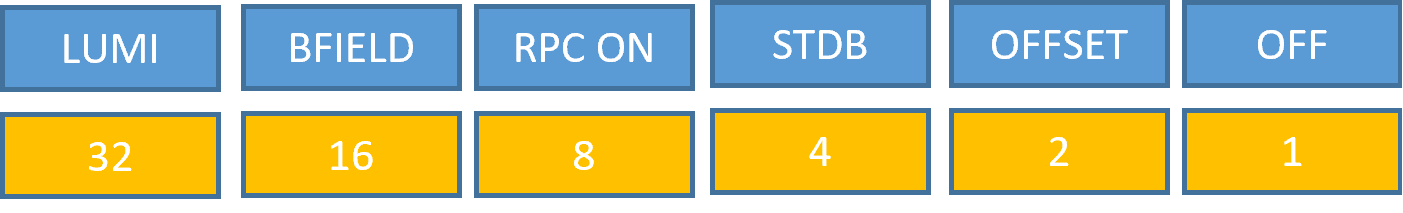}
\caption{ \label{fig:tag} Six binary bits are used to encode additional information for each RPC HV channel current. The rightmost four bits represent the HV state and voltage range for each channel, while the two leftmost bits encode critical information about the CMS magnetic field strength and the presence of colliding beams with non-zero instantaneous luminosity. }
\end{figure}

Data tagging is essential for ensuring the accuracy and relevance of analyses within the RPC automation framework. This process associates every current entry in the \textit{RPCCURRENTS} table with specific operational conditions, including high HV states, the presence of the CMS magnetic field, and LHC beam collisions. A dedicated column in the \textit{RPCCURRENTS} table, shown in Fig. \ref{fig:flag}, stores the FLAG data, which encodes these conditions as a decimal representation of six binary condition bits (values from 0 to 63), see Fig. \ref{fig:tag}.

The least significant bit, OFF, indicates whether the HV channel is turned off (value 0 when ON). The second bit, OFFSET, tags currents recorded when the voltage is set to 1000 V. The third bit, STDB, identifies currents at the standard standby voltage for the detector. The fourth bit, RPC ON, marks currents in the gas multiplication regime, reflecting the operational state of the RPC detectors. These HV condition flags are derived internally from the synchronized \textit{RPCCURRENTS} table within the \textit{CMS\_RPC\_COND} schema.

The remaining two bits are sourced from external schemas. The CMS magnetic field data is retrieved from the \textit{CMSFWMAGNET} table of the \textit{CMS\_DCS\_ENV\_PVSS\_COND} schema, with the BFIELD bit set to 1 when the magnetic field exceeds 1.99 T. The most significant bit, LUMI, indicates the presence of beam collisions, signaling active luminosity conditions during data acquisition, and is retrieved from the \textit{LUMISECTIONS} table in the \textit{CMS\_OMS} schema \cite{cms_oms}. Together, these tags provide a robust framework for filtering and analyzing currents under well-defined operational conditions.
\subsection{\label{subsec:block} Block Averaging Method}
In many cases, an RPC HV channel is maintained at a fixed voltage for a certain period, with external conditions also held constant during this time. In such instances, we may want to represent the entire period with a single current value. 

We refer to this period as a block, and the current value for the block is calculated by averaging the current over the duration of the block, following specific criteria. 
\begin{figure}[htbp!]
\includegraphics[width=0.99\columnwidth]{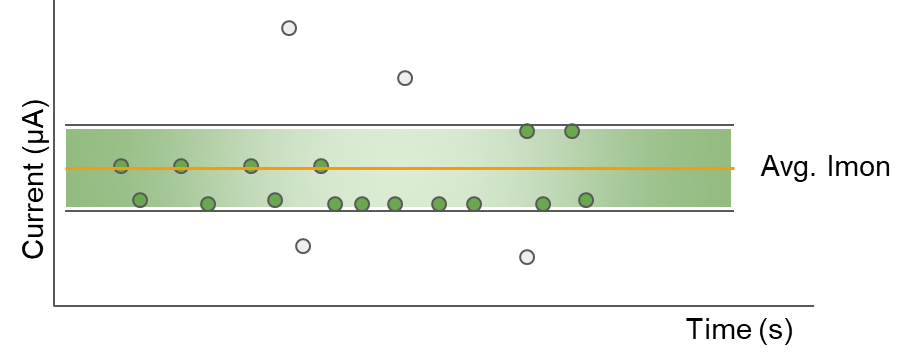}
\caption{ \label{fig:block} The average block current is calculated by iteratively removing outliers (outside ±4$\sigma$) from the Imon distribution until no more than 10\% of the data points are discarded, then computing the mean for the period. This method is used in various automata in the framework to provide the average current per block. }
\end{figure}

First, we retain data points from the current (Imon) distribution that lie within ±4$\sigma$ of the mean, discarding any points outside this range. This process is applied iteratively, removing outliers until no more than 10\% of the original data points are excluded. The resulting mean value is then considered the "average block current" for that period, Fig. \ref{fig:block}. 

This method is employed to define a single current value for offset, cosmic, standby, HV conditioning blocks, and initial ramp-up blocks, which are discussed in detail in subsequent sections \ref{subsec:hvc}, \ref{subsec:om}, and \ref{subsec:evo}.
\subsection{Data Streaming and Multithreading Techniques}
Data streaming and multithreading are essential techniques for handling large volumes of real-time data efficiently. Data streaming allows continuous ingestion and processing of data in small, manageable chunks, enabling real-time analysis without overwhelming system resources. This approach is particularly useful in scenarios where data arrives at high velocity, such as in particle physics experiments. Multithreading, on the other hand, enables parallel execution of tasks, maximizing CPU utilization and reducing processing time. By distributing tasks across multiple threads, multithreading ensures that the system can handle multiple processes concurrently, improving performance and responsiveness in time-sensitive applications.

Both techniques are crucial for optimizing data handling in the RPC automation framework, enabling fast, efficient, and scalable processing of real-time data from the RPC detectors. 
\subsection{\label{subsec:lhcb} LHC Block Concept}
LHC machine modes, such as Proton Physics, Ion Physics, Machine Development, and Shutdown, define the overall operational state of the LHC \cite{lhc}, varying based on the type of experiment or maintenance being conducted. Beam modes, on the other hand, specify the phases of beam preparation, acceleration, collision, and dump during LHC operation.

During the Ion Physics machine mode, RPC currents are very low and nearly identical to cosmic currents when no beam is present. As a result, LHC blocks in the RPC automation framework are only defined for the primary machine mode of operation, Proton Physics.

For RPC current data analysis within the automation framework, four key LHC blocks are defined: Stable Beams, Cosmic, Standby, and Cool-down. The Stable Beams block corresponds to the beam collision period when physics data is collected, typically lasting from the Stable Beams to Beam Dump LHC beam modes. The cosmic block occurs between the Beam Dump LHC beam mode and the Inject Warning signal of the Injection handshake, marking the period when no beam is present in the machine. The Cool-down block is a subset of the cosmic block, situated at its start, between the Beam Dump and Ramp Down machine modes. Lastly, the standby block spans from the Inject Warning to Squeeze, during which the LHC is engaged in beam injection and energy ramping.
\begin{figure}[htbp!]
\includegraphics[width=0.99\columnwidth]{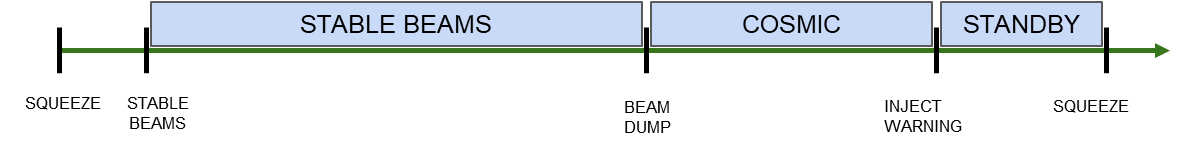}
\caption{ \label{fig:lhc blocks} The four key LHC blocks for RPC current data analysis are Stable Beams, Cosmic, Standby, and Cool-down, each corresponding to distinct operational periods in the LHC beam cycle. The Cool-down block is not depicted in the plot, as it is too short (lasting only a few seconds) to be effectively visualized. }
\end{figure}

The definition of LHC blocks is presented in Fig. \ref{fig:lhc blocks}, and the use of these blocks by various framework automata is detailed below.

During the Stable Beams period, we conduct a current vs luminosity study and integrate the collision currents to calculate the integrated charge during collisions. In the cosmic gap, we apply the block averaging method (see section \ref{subsec:block}) to average the cosmic currents for each HV channel within each LHC cosmic block, tracking their evolution over time and integrating them to determine the integrated charge outside of collisions. Similarly, during the standby gap, we use the block averaging method (see section \ref{subsec:block}) to average the standby currents for each HV channel within each LHC standby block, allowing us to monitor their evolution over time.
\subsection{\label{subsec:lumimethod} LHC Luminosity Method}
LHC instantaneous luminosity is a crucial parameter for analyzing RPC currents and rates. Initially, studies compared online and offline luminosity data, applying a dedicated lumi-filter to align the online luminosity with its refined offline counterpart. However, these early analyses were run-based, averaging currents and luminosities across entire data-taking runs, which diluted precision and excluded peak instantaneous luminosities from consideration. Furthermore, robust correlation studies required data from about 1000 runs spanning an entire year of operation.
\begin{figure}[htbp!]
\includegraphics[width=0.99\columnwidth]{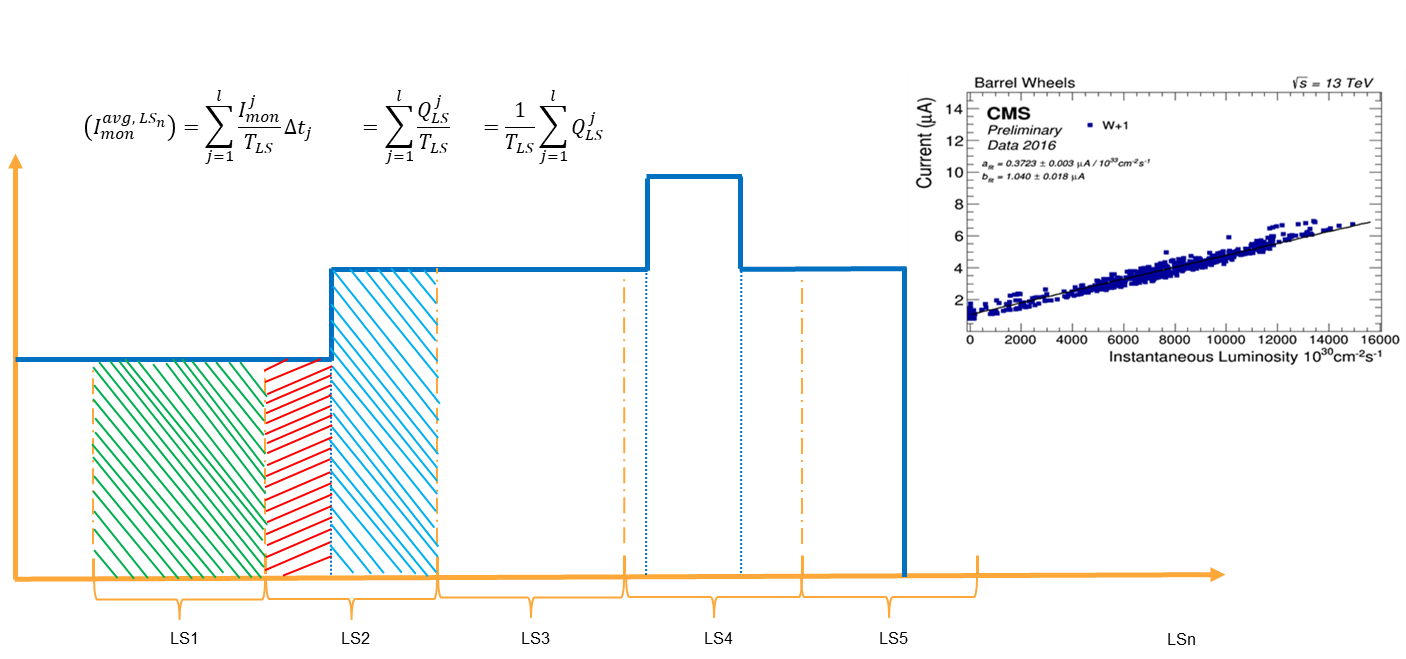}
\caption{ \label{fig:lumi4} The current is integrated over lumisections to calculate the charge per lumisection, which, when divided by the lumisection duration, yields the average current per lumisection. This value is assigned as the current per lumisection. Applying a linear fit to all current vs. instantaneous luminosity points per fill for a given HV channel or a bigger granularity object, enables accurate determination of the current dependency immediately at the end of each fill.}
\end{figure}

These limitations are addressed by implementing a lumisection-based method, where currents are averaged per lumisection, a data-taking interval of approximately 23.01 seconds. Since the concept of  current per lumisection did not exist in database, a novel streaming technique was developed, as illustrated in Fig. \ref{fig:lumi4} to assign a current value per lumisection for each HV channel. This approach allows the collection of 1000 data points within a single LHC fill rather than over a year. By analyzing lumisection data instead of run-based data, the studies now incorporate high instantaneous luminosity values. 

This innovative method, implemented through the RPC automation framework, provides immediate results at the end of each LHC fill, significantly enhancing the efficiency and timeliness of data analysis.
\section{\label{idea} Ideology}
\begin{figure}[htbp!]
\centering
\includegraphics[width=0.99\columnwidth]{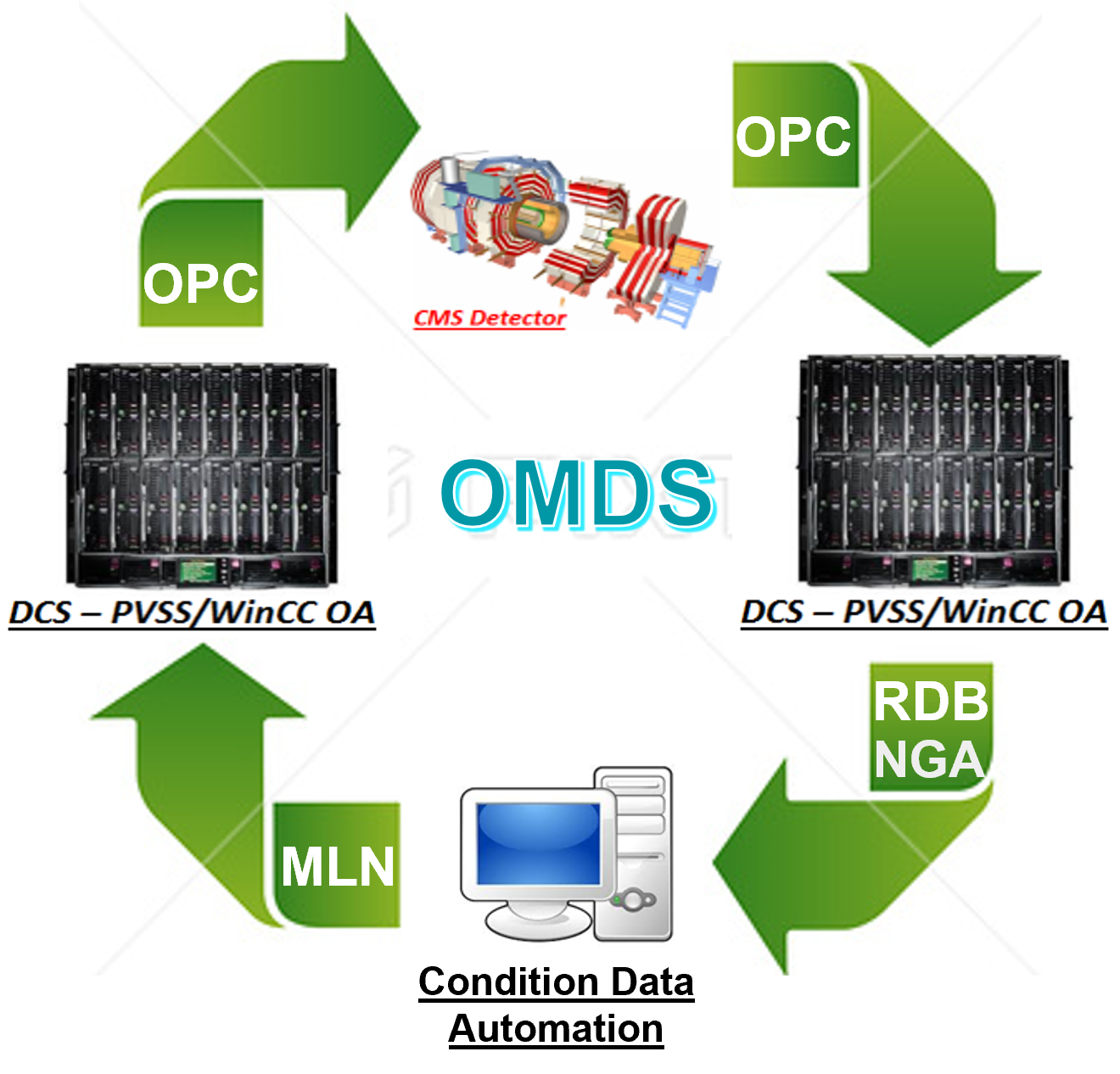}
\caption{ \label{fig:fullcircle} The project’s ideology involves synchronizing condition data, performing analytical studies (right flow), refining their results through the Machine Learning Navigator (MLN) for predictive modeling, and automatically adjusting detector parameters via OPC drivers to ensure self-monitoring and self-correcting operations (left flow).}
\end{figure}

The core ideology of the project, within which RPC automation plays a key role, is depicted in Fig. \ref{fig:fullcircle}. 

All condition data, complementing standard physics data-taking, is retrieved from detector hardware channels and transmitted via the OPC\footnote{OPC stands for OLE for Process Control, where OLE means Object Linking and Embedding. OPC UA (Unified Architecture) is widely used in industrial and process control systems for secure and reliable communication.} protocol to the non-physics event bus. This data, previously archived by the Raima Database (RDB) manager, is now managed by the CMS Next Generation Archiver (NGA) and stored in the \textit{OMDS} database on production \cite{nga} - Fig. \ref{fig:fullcircle} (right).

The condition data automation, introduced in section \ref{sec:impl}, synchronizes raw data, processes it for analytical studies, and records the analyzed results in a dedicated schema on the production database. From there, the Machine Learning Navigator (MLN) integrates the refined data into predictive models. These models forecast current behavior and issue notifications on deviations over a threshold between measured and predicted currents. The final phase of the cycle involves automated algorithms processing these warnings and errors, taking decisions, and transmitting updated commands via OPC drivers to the detector to adjust parameters as needed - Fig. \ref{fig:fullcircle} (left).

This closed-loop system represents the full-circle ideology of self-monitoring and self-correcting detector operations.
\section{Main Automata}
Main automata are core components of the RPC automation framework, responsible for synchronizing raw detector data and structuring it for comprehensive analysis and storage.
\subsection{UXC Environment}
\begin{figure}[htbp!]
\includegraphics[width=0.99\columnwidth]{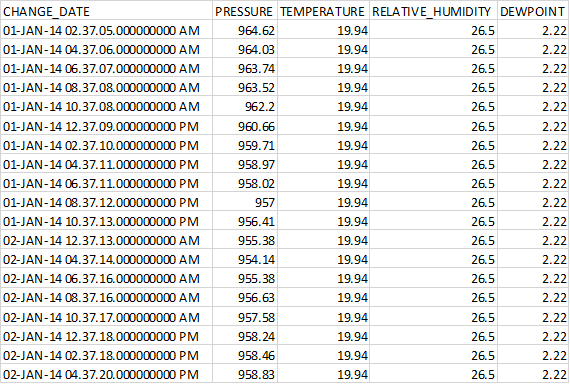}
\caption{ \label{fig:envpar} The \textit{UXC\_Environment} automaton retrieves and synchronizes CMS cavern pressure, temperature, relative humidity, and dew point data.}
\end{figure}

The first automaton in the RPC framework automation chain, \textit{UXC\_Environment}, retrieves raw data on pressure, temperature, relative humidity, and dew point within the CMS experimental cavern (UXC) and records it synchronously in a dedicated table within the \textit{CMS\_RPC\_COND} schema which format is shown on Fig. \ref{fig:envpar}.
\subsection{LHCLOG}
The \textit{LHCLOG} automaton systematically extracts timestamps of machine, beam, and handshake mode changes from the \textit{CMS\_DCS\_ENV\_PVSS\_COND} schema using regular expressions to identify predefined patterns. The extracted data is stored in an optimized database table, providing a comprehensive record of these changes. Complementary automata define and record LHC blocks for Stable Beams, cosmic gaps, standby, and Cool-down periods, as detailed in Section \ref{subsec:block}.
\subsection{RPCCURRENTS}
The main automaton in the entire RPC condition data automation framework is the \textit{RPCCURRENTS}, which retrieves all raw currents from the \textit{CMS\_RPC\_PVSS\_COND} schema, synchronizes them, and tags them with RPC HV, CMS magnetic field, and LHC instantaneous luminosity information (see Section \ref{subsec:tagging}) into the RPC automation schema on \textit{OMDS}.
\begin{figure}[htbp!]
\includegraphics[width=0.99\columnwidth]{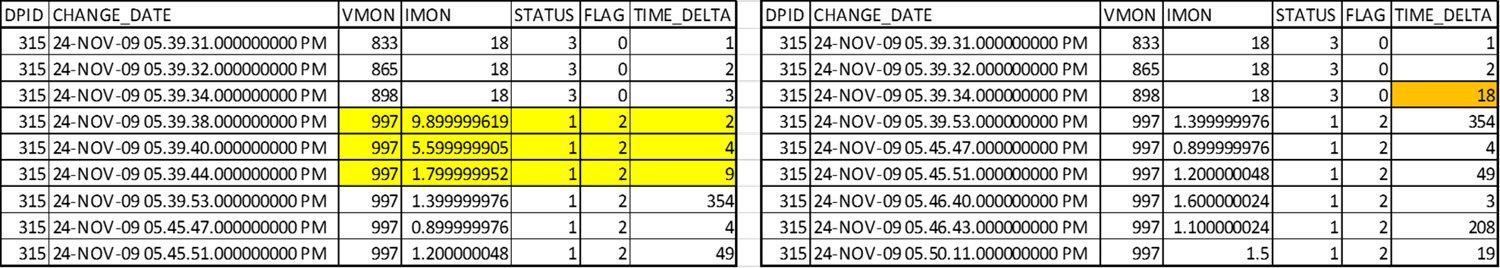}
\caption{ \label{fig:rUp} Filtering out current decay, highlighted in yellow, due to the capacitive detector effect after voltage ramp-up (left) ensures stable status "1" data by reassigning removed rows to the ramp-up phase, combining them with the last status "3" entry, highlighted in orange (right), to preserve timeline continuity.}
\end{figure}

Furthermore, two additional automata are executed sequentially after the main automaton to perform current data smoothing, which involves removing ramp-up and ramp-down currents at stable monitored voltages.

Due to their large surface area, RPC behave as large capacitors, leading to a significant current decay following the voltage ramp-up phase. While the hardware channel status immediately switches to "1" to indicate stable voltage, the current does not stabilize as quickly due to the chambers' capacitive effect (see Fig. \ref{fig:rUp}, left). This transient behavior results in artificial cosmic current spikes, which motivated one of the project's initial goals: cleaning such anomalies.

The ramp-up automaton addresses this issue by employing a method similar to the block averaging approach described in Section \ref{subsec:block}. It identifies stable current behavior and removes data points outside the defined stability bandwidth. To avoid introducing gaps in the timeline, filtered current decay rows are reassigned to the ramp-up phase and tagged with status "3" (see Fig. \ref{fig:rUp}, right). This ensures accurate representation of current behavior without disrupting temporal continuity.

\begin{figure}[htbp!]
\includegraphics[width=0.99\columnwidth]{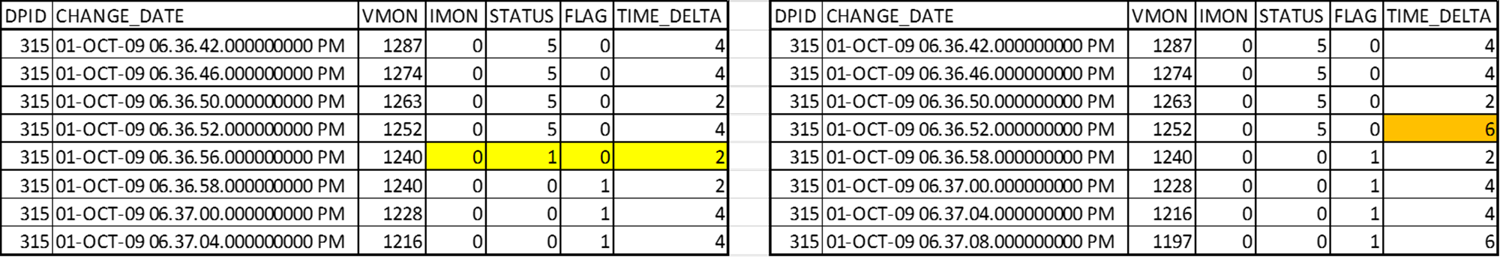}
\caption{ \label{fig:rDw} Filtering out erroneous status "1" during ramp-down, highlighted in yellow (left), ensures a direct transition from status "5" (ramp-down) to status "0" (off), with the removed row’s duration added to the last status "5" entry, highlighted in orange (right), preserving timeline continuity.}
\end{figure}

The ramp-down automaton eliminates erroneous status "1" entries generated by the hardware firmware during the ramp-down phase, enabling a direct transition from status "5" (ramp-down) to status "0" (off), as highlighted in Fig. \ref{fig:rDw}, left. The duration of the removed status "1" row is appended to the final status "5" entry, Fig. \ref{fig:rDw}, right, to maintain timeline continuity.
\subsubsection{\label{subsec:vircur} Virtual Currents}
The introduction of virtual objects in the RPC automation framework marked a significant advancement, enabling data analysis and storage for lower-granularity objects. These include Barrel and Endcap regions as a whole, Barrel wheels, stations, and sectors, as well as Endcap disks, rings, and sectors. This innovation facilitates the study of current distributions along the CMS beampipe axis (\textit{z}), radially outward from the beampipe (\textit{R}), and azimuthally ($\phi$) in the CMS coordinate system.

Virtual objects are assigned unique data point identifiers in the \textit{CMS\_COND\_RPC} schema on the \textit{OMDS} instance of the production database, despite having no direct correspondence to hardware channels. Data for these virtual objects is aggregated based on the detector's geometrical sections, as defined by a DPID map. The aggregated averaged current (Imon) value is recorded for each virtual DPID at every timestamp when the Imon value of any associated hardware DPID changes. Virtual currents are stored and utilized for further studies in the same manner as standard HV channel currents.
\subsection{RPCRATES}
\begin{figure}[htbp!]
\includegraphics[width=0.99\columnwidth]{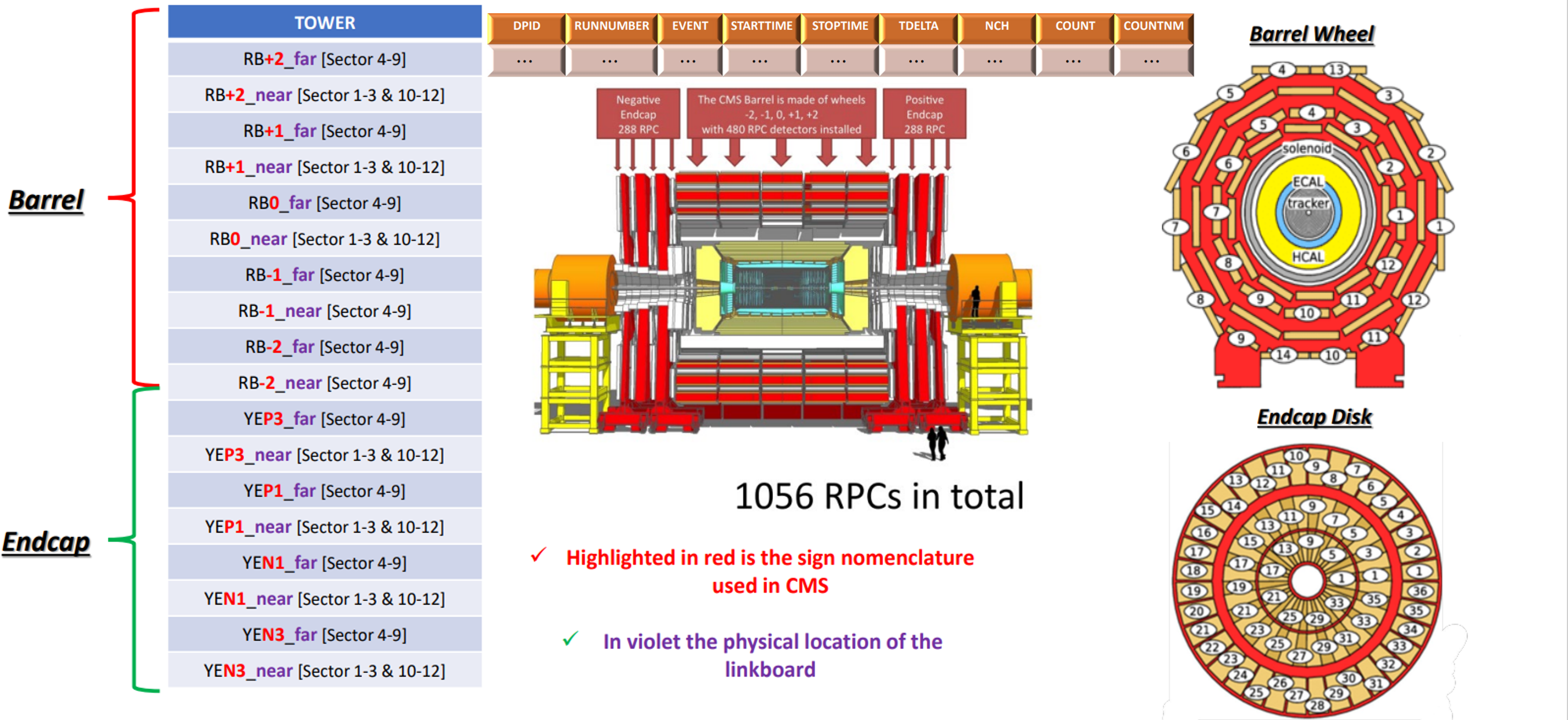}
\caption{ \label{fig:rate} Raw data from 2,748 link boards, collected from 1,056 Resistive Plate Chambers operated in Run-II and Run-III, is transferred from 18 \textit{ROOT} files per run into a structured database table format in the RPC automation schema on \textit{OMDS}, as shown in the top middle section of the figure. The physical locations of the trigger towers, indicating the sectors from which data is collected, are also presented.}
\end{figure}

RPC trigger rates \cite{rpc_trigger} are the second major non-physics event parameter used in RPC correlational studies, following currents. Unlike currents, rate raw data is stored in \textit{ROOT} \cite{root} files rather than database tables. The \textit{RPCRATES} automaton opens these \textit{ROOT} files to extract the necessary data and stores it in the RPC automation schema on \textit{OMDS}, as shown in Fig. \ref{fig:rate}.

Raw data files are stored per run number and trigger tower, with 18 RPC trigger towers—10 in the Barrel and 8 in the Endcap regions—resulting in 18 \textit{ROOT} files per run. Unlike currents, which are continuously monitored, trigger rates are recorded only during data-taking when the RPC system is included either in the central Data Acquisition system (DAQ) \cite{cms_daq} or is running locally. Thus, a key distinction is that RPC trigger rates are recorded on a per-run basis, rather than continuously over time. Trigger rates, like currents, can be categorized into collision and cosmic rates, though present studies focus solely on collision rates, without tagging for LHC beam or CMS magnetic field presence.

The \textit{ROOT} files contain the number of counts within a given readout time window, referred to as an "event" in the database table format, which length can vary, and is defined by the mutex operator \cite{rpc_trigger}. This operator determines when a given link board (LB), part of the RPC link system \cite{rpc_link_upgrade} connecting the front-end electronics with the trigger electronics in the CMS underground service cavern (USC), is read out. The database records the start and stop times of each event in a run, along with the counts per LB per event. However, trigger rates are not directly stored in the database but are instead calculated during data streaming by dividing the counts by the event duration in seconds, a process handled by the study automata.
\subsubsection{Virtual Rates}
Virtual trigger rates extend the concept of virtual currents, described in Section \ref{subsec:vircur}, to lower-granularity correlational studies within the RPC automation framework. These rates are linked to virtual objects, each identified by a unique data point identifier, and aggregated across link boards according to a predefined DPID map. This enables the analysis of rates across different detector segments, such as the Barrel and Endcap regions, and their respective wheels, disks, stations, rings, and sectors.

Averaged rate values for each virtual DPID are recorded whenever any associated hardware DPID triggers a change, allowing for detailed studies of trigger rates in various detector segments. This aggregation offers valuable insights into RPC performance, facilitating trend analysis across the detector.
\subsection{RPC Gas Channel Flow}
\begin{figure}[htbp!]
\includegraphics[width=0.99\columnwidth]{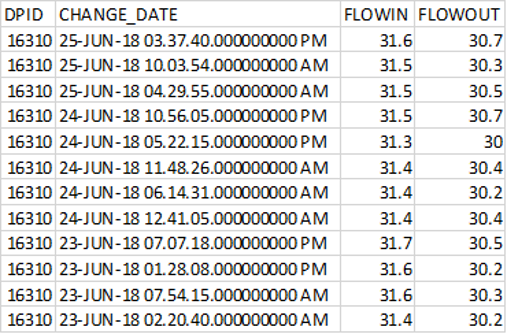}
\caption{ \label{fig:gasflow} \textit{RPCGASFLOW} automaton synchronizes raw data for the main gas parameters, flow-in and flow-out rates, and stores it in the \textit{RPCGASFLOW} table in the \textit{CMS\_RPC\_COND} schema.}
\end{figure}

The last automaton in the automation chain, \textit{RPCGASFLOW}, retrieves asynchronous raw data for gas flow-in and flow-out rates from the \textit{CMS\_RPC\_PVSS\_COND} schema and synchronizes them as described in Section \ref{subsec:synch} into a dedicated table in the RPC automation schema on \textit{OMDS} for further analysis, see Fig. \ref{fig:gasflow}.
\section{Auxiliary Automata}
After the main automata complete raw data synchronization and storage in a structured database format, the auxiliary automata perform the primary correlational studies of fundamental RPC detector characteristics. These include examining currents and rates in relation to environmental conditions (e.g., relative humidity), operational parameters (e.g., gas flow rate), or LHC beam properties (e.g., instantaneous luminosity).

In addition to correlational analysis, certain auxiliary automata focus on RPC detector specific studies, such as Integrated Charge, Current Evolution, Operation Mode Identification, HV Conditioning, and Active Channels, providing deeper insights into detector behavior and performance.
\subsection{\label{subsec:ic} Integrated Charge}
The RPC Integrated Charge automaton, first developed in 2017, is designed to aid aging studies and estimate the charge to be integrated at 3000 $fb^{-1}$ of LHC integrated luminosity \cite{rpcic}. This estimation, scaled by a safety factor of 3, has since guided irradiation tests at the CERN GIF++ facility \cite{gifpp}.

\begin{figure}[htbp!]
\includegraphics[width=0.99\columnwidth]{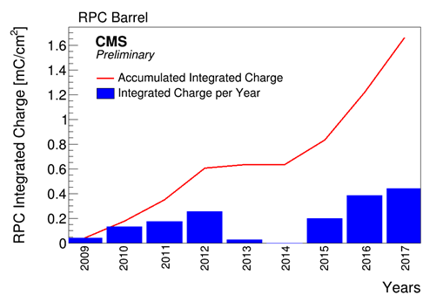}
\caption{ \label{fig:ic} Daily integrated charge values per RPC HV channel and virtual objects, computed for \textit{collision} and \textit{RPC\_ON} currents, are recorded in the \textit{CMS\_RPC\_COND} schema on production database. Graphs display accumulated and yearly integrated charge for the Barrel region as a virtual object at the end of 2017, demonstrating an example of the automation's utility.}
\end{figure}

For each HV channel (HV DPID), the automaton computes the charge by integrating currents over the time intervals between adjacent records in the \textit{RPCCURRENTS} database table, yielding daily integrated charge values. Two types of daily values are provided per HV channel: \textit{Collision} and \textit{RPC\_ON}. The \textit{Collision} type integrates only collision currents, tagged with Flag 56 as per Section \ref{subsec:tagging}, while the \textit{RPC\_ON} type considers all currents when HV channels are set to working points, thus encompassing both collision and cosmic currents. This computation spans the entire RPC operational history, starting in October 2009. As generally valid for all auxiliary automata, the RPC Integrated charge is also available for virtual objects as presented in Fig. \ref{fig:ic}.
\subsubsection{\label{subsec:aic_dpid} Accumulated Integrated Charge per HV channel}
\begin{figure}[htbp!]
\includegraphics[width=0.99\columnwidth]{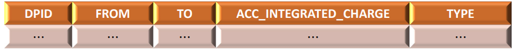}
\caption{ \label{fig:aic_dpid} Accumulated integrated charge per HV channel and virtual objects, calculated daily since 2009, is stored with a consistent \textit{FROM} date and an incrementing \textit{TO} date, providing a cumulative charge value for each day of operations.}
\end{figure}

The RPC Accumulated Integrated Charge automaton consolidates daily integrated charge values from the Integrated Charge automaton described in Section \ref{subsec:ic} into a dedicated database table, which structure is presented in Fig. \ref{fig:aic_dpid}. Additionally, it records a single accumulated charge value per day for both HV channels and virtual objects, spanning from the start of operations in October 2009, see Fig. \ref{fig:ic}. This enables fast, real-time plotting of accumulated values via graphical interfaces and facilitates correlation analyses using accumulated charge as a proxy for time.
\subsubsection{Accumulated Integrated Charge per Chamber}
\begin{figure}[htbp!]
\includegraphics[width=0.99\columnwidth]{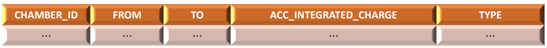}
\caption{ \label{fig:aic_chamber} The RPC HV channel reshuffling in October 2014 resulted in \textit{DPID} reassignment, requiring the adoption of unique \textit{Chamber\_IDs} for accurate and efficient processing of accumulated integrated charge data. A new database structure, based on \textit{Chamber\_IDs}, was created alongside the original \textit{DPID}-based structure in Fig. \ref{fig:aic_dpid}.}
\end{figure}

To address efficiency optimization in the Endcap region, where a single HV channel powers two chambers, a HV channel reshuffling was implemented in October 2014, necessitating DPID reassignment. For parameters like integrated and accumulated integrated charge, whose accurate calculation spans the entire operational timeline, a transition from HV channel DPID to a unique ChamberID was essential, Fig. \ref{fig:aic_chamber}. This paradigm change ensures correct and efficient data processing and storage, attributing charge data to the appropriate chamber entity.
\subsection{\label{subsec:hvc} HV Conditioning}
RPC HV Conditioning consists of periodic current scans conducted 3-4 times annually at predefined voltage steps: 1000, 2000, 3000, 4000, 5000, 6000, 7000, 8000, 8500, 9000, 9100, 9200, 9300, 9400, 9500, and 9600 V.

Each HV point represents a block of current data recorded at a stable voltage. The block averaging method described in Section \ref{subsec:block} is used to derive a representative current value for each block. These values are compiled into a complete conditioning set encompassing the full sequence of HV points and are stored separately in the database, enabling efficient and comprehensive plotting of HV conditioning data on graphical interfaces.
\subsection{HV Conditioning Fit}
The RPC HV Conditioning Fit automaton processes HV points between 1000 and 7000 V from each conditioning data set, for both HV channels and virtual objects. 
\begin{figure}[htbp!]
\includegraphics[width=0.99\columnwidth]{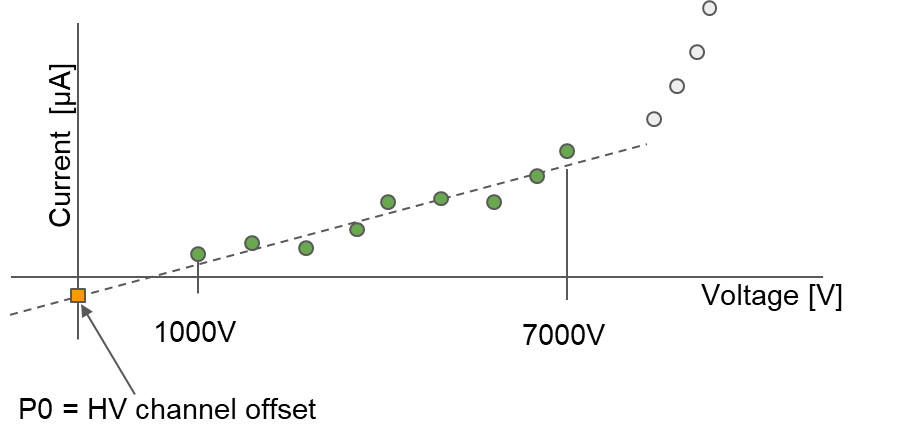}
\caption{ \label{fig:hvcond} Linear fits in the ohmic region (1000–7000 V) of RPC HV conditioning sets yield fit parameters $p_0$, HV channel offset, and $p_1$,  inverse resistance, stored for analysis.}
\end{figure}

A linear fit is performed on the conditioning subset of data in the ohmic region as depicted in Fig. \ref{fig:hvcond}. Fit optimization is performed during data streaming, with the final results stored in a dedicated database table \textit{RPCCURRENTS\_HVCOND\_FIT}. The stored parameters include the fit coefficients: $p_0$ (HV board channel offset) and $p_1$  (inverse resistance), allowing fast data retrieval for plotting of Resistance (1/$p_1$) over time.
\subsection{\label{subsec:om} Operation Mode}
\begin{figure}[htbp!]
\includegraphics[width=0.99\columnwidth]{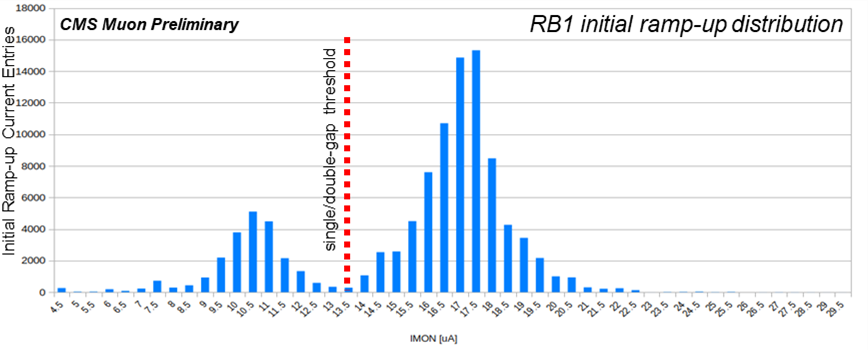}
\caption{ \label{fig:om} The RPC operation mode automaton determines whether a chamber operates in single-gap or double-gap mode based on initial ramp-up current and a predefined threshold per chamber type between 500V and 1030V.}
\end{figure}

The RPC Operation Mode automaton determines whether RPC chambers operate in single-gap (SG) or double-gap mode when powered. Each RPC chamber comprises two HV layers: Up and Down. In \textit{DG} mode, both layers are operational, and signals induced by both gaps are collected onto the central readout plane, maximizing efficiency. If one HV layer is unpowered, the dark currents during the ramp-up phase are approximately halved, providing a critical indicator of \textit{SG} operation mode, Fig. \ref{fig:om}.

The amplitude of the ramp-up current is directly influenced by the gap surface area: larger gaps draw higher currents. To account for this, specific thresholds are defined for each chamber type, including RB1, RB2/2, RB2/3, RB3, RB4-1500, RB4-2000, RB4-2500, RE1/2, RE1/3, and REn/2 or REn/3, where n denotes the station number and $n \in [2,4]$. These thresholds are applied during the initial ramp-up phase, between 500 V and 1030 V, to determine the operation mode of each detector.
\subsection{Lumi Fit}
The most significant study requiring advanced automation is the dependence of key RPC characteristics, such as currents and rates, on the LHC instantaneous luminosity. The Lumi Fit automata analyze these correlations, offering valuable insights into detector performance under varying beam conditions.

The LHC life-cycle includes distinct phases—Stable Beams, cosmic gaps, and standby gaps—utilized for various studies of key RPC parameters. Collision currents during Stable Beams are analyzed for current vs. luminosity dependence, while cosmic and standby gaps are used to monitor the stability and evolution of cosmic and standby currents (Fig. \ref{fig:lumievo}).
\subsubsection{Current Lumi Fit}
Instantaneous luminosity values, directly measured by luminometers and provided via the non-physics event bus, are paired with collision-tagged currents for each HV channel or virtual object. As detailed in Section \ref{subsec:lumimethod}, average collision currents are assigned to lumisections, enabling precise alignment with luminosity data for correlation studies.
\begin{figure}[htbp!]
\includegraphics[width=0.99\columnwidth]{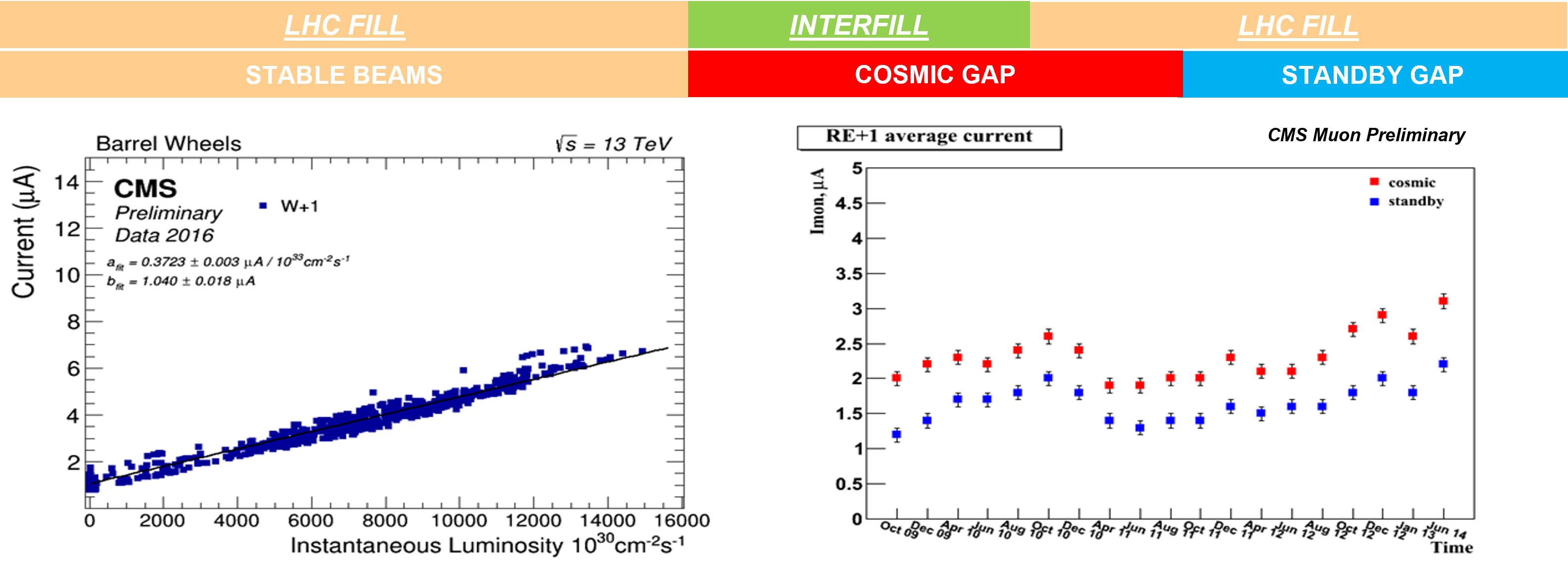}
\caption{ \label{fig:lumievo} LHC life-cycle phases—Stable Beams, cosmic gaps, and standby gaps—are used to study RPC current dependence on luminosity and monitor cosmic and standby current stability over time. Nearly 900 current vs. luminosity fits per fill align collision-tagged currents with instantaneous luminosity, enabling studies of detector performance, shielding effects, and stability across beam conditions (left).}
\end{figure}

A linear fit is applied to the data, with the intercept representing the beam-dump current (current at zero luminosity) and the slope indicating the current's dependence on luminosity. These fit parameters are stored in the database, supporting further analysis of detector performance, the influence of neutron shielding, and long-term stability across different luminosity levels. Close to 900 current vs. luminosity fits are generated for each LHC fill, Fig. \ref{fig:lumievo} (left), encompassing all HV channels and virtual objects, providing comprehensive insight into the detector's response under varying beam conditions.
\subsubsection{Rate Lumi Fit}
Similarly, RPC trigger rates for each object are assigned to lumisections using the methodology described in Section \ref{subsec:lumimethod}. Average rates are calculated per lumisection and matched with the corresponding luminosity values.

\begin{figure}[htbp!]
\includegraphics[width=0.99\columnwidth]{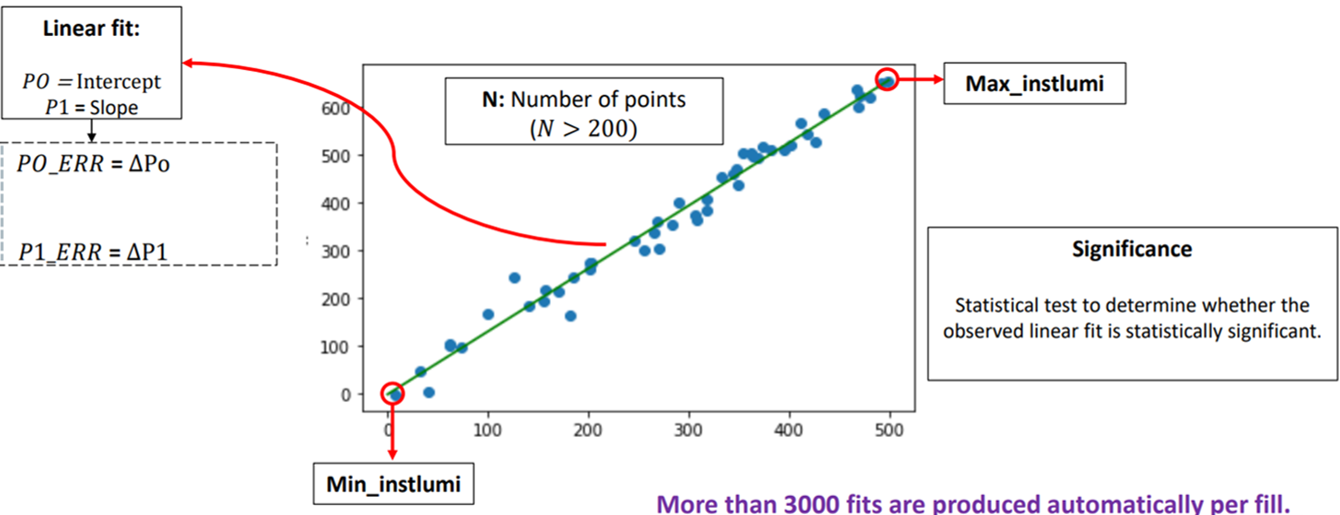}
\caption{ \label{fig:ratelumi} Over 3000 rate vs. luminosity fits per fill analyze RPC trigger rates across link boards and virtual objects vs. LHC instantaneous luminosity, revealing detector behavior in high-luminosity conditions.}
\end{figure}

A linear fit is applied, where the intercept represents the cosmic rate (rate at zero luminosity), and the slope quantifies the rate's increase per unit of instantaneous luminosity. Over 3000 rate vs. luminosity fits, similar to the one shown on Fig. \ref{fig:ratelumi}, are generated per LHC fill, covering all link boards from all trigger towers and virtual objects, enabling detailed analysis of the detector's response across varying luminosity scenarios. These results are recorded in the database for ongoing performance assessments, deepening the understanding of RPC detector behavior during high-luminosity operations.
\subsection{\label{subsec:evo} Evolution}
One of the key auxiliary automata focuses on analyzing the evolution of RPC currents over time. Changes in current behavior can have significant implications for detector performance, making it critical to validate their causes and potential effects. To enable detailed studies, three distinct current types are defined: offset, standby, and cosmic currents.

\begin{figure}[htbp!]
\includegraphics[width=0.99\columnwidth]{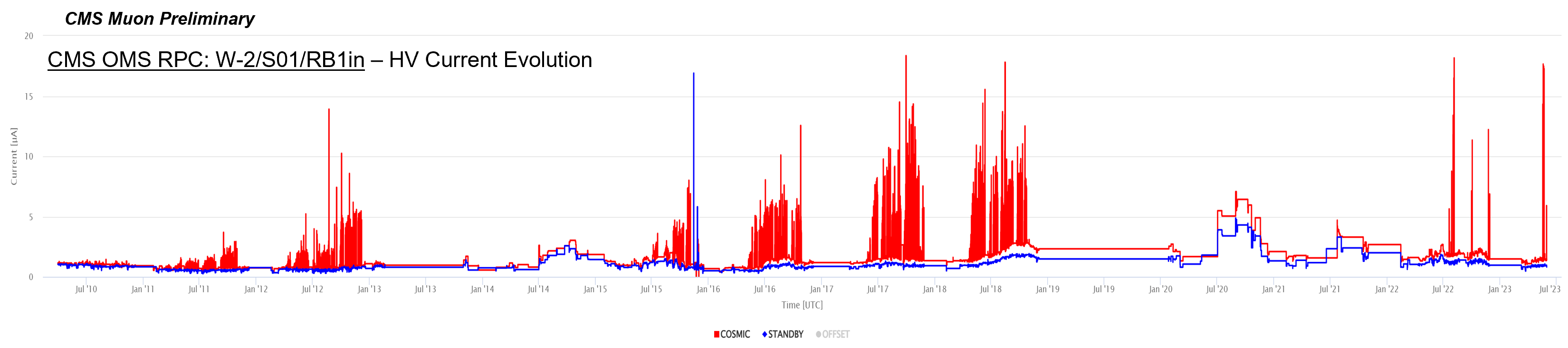}
\caption{ \label{fig:evo} Current evolution studies focus on offset, standby, and cosmic currents. Using the block averaging method, single mean values per block are computed and stored, significantly reducing data volume while preserving key information for trend analysis as shown for one chamber, \textit{W-2/S01/RB1in}, in the Barrel region.}
\end{figure}

Given the vast volume of raw data accumulated per HV channel since the start of CMS operations, efficient data reduction is necessary. The block averaging method, detailed in Section \ref{subsec:block}, plays a pivotal role in this process. Current blocks are categorized as offset (measured at 1000 V), standby (measured at 6500 V), or cosmic (measured at working points during beam-off periods). For each block, a single mean value is computed and stored in the database, provided constant conditions are maintained throughout the block's duration. This approach ensures manageable data volumes while retaining the essential information for trend analysis. In Fig. \ref{fig:evo}, the current evolution for a single Barrel chamber is presented. Noticeable are numerous false cosmic-tagged current spikes, caused by fluctuations in instantaneous or delivered luminosity. These instabilities, linked to Van-der-Meer scans or luminosity leveling, result in incorrect tagging of collision currents as cosmic.
\begin{figure}[htbp!]
\includegraphics[width=0.99\columnwidth]{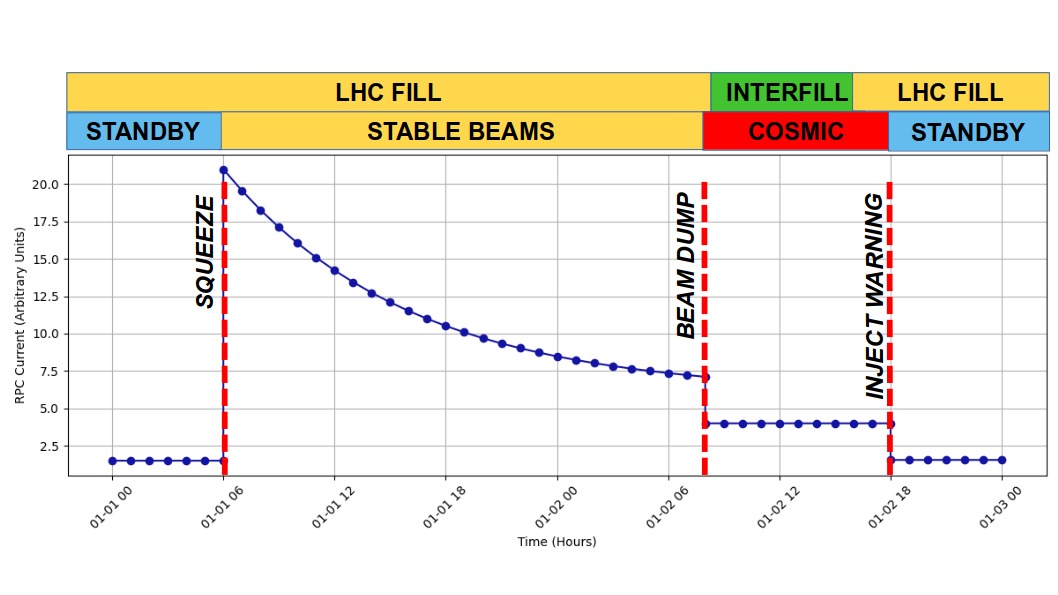}
\caption{ \label{fig:curblocks} Typical RPC collision, cosmic, and standby current behavior during LHC fills and interfill gaps. Cosmic blocks form during no-beam cosmic gaps, while standby blocks form during injection and energy ramp phases of the LHC fill.}
\end{figure}

\begin{figure}[htbp!]
\includegraphics[width=0.99\columnwidth]{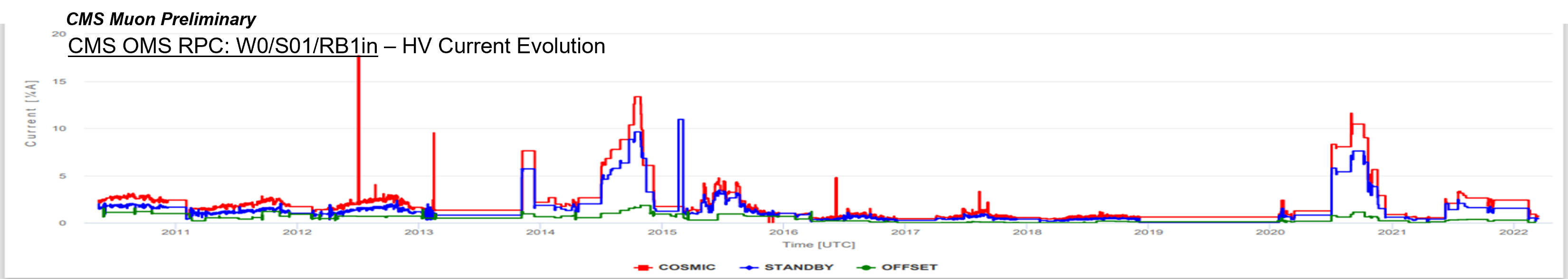}
\caption{ \label{fig:evoclean} Example of cleaned current evolution for a Barrel chamber, \textit{W0/S01/RB1in}, aligned with LHC blocks to eliminate erroneous spikes.}
\end{figure}

To address this issue, standby, cosmic, and Stable Beams current blocks are matched with corresponding LHC blocks: standby gap, cosmic gap, and Stable Beams periods, as depicted in Fig. \ref{fig:curblocks}. By combining the concept of current blocks from Section \ref{subsec:block} with LHC blocks described in Section \ref{subsec:lhcb}, only current blocks aligned with LHC blocks are designated as clean data, ensuring reliable plotting and analysis. Figure \ref{fig:evoclean} illustrates the cleaned evolution of a Barrel chamber, demonstrating the removal of erroneous spikes by considering only cosmic current blocks aligned with LHC cosmic gaps for cosmic current evolution. Similarly, only standby current blocks matching LHC standby gaps are used for the standby current evolution of RPC chambers.

In addition to the three current evolution automata for offset, standby, and cosmic current trend analysis, additional automata are implemented to monitor initial ramp-up currents, HV trips, and over-current spikes. Together, these automata provide a comprehensive view of RPC operational performance.

The \textit{Initial Ramp-up Currents} automaton applies the block averaging method (Section \ref{subsec:block}) to calculate the mean current values during the detector's initial ramp-up phase, between 500 and 1030 V. These averaged ramp-up currents are subsequently utilized by the \textit{Operation Mode} automaton (detailed in Section \ref{subsec:om}) to determine the mode of operation for each detector.

Two dedicated automata handle HV channel trips, distinguishing between operation trips and ramp-up trips. Leveraging the framework's block methods, the \textit{Operation Trip} automaton identifies sequences of blocks with CAEN hardware statuses 1, 9, or 512, while the \textit{Ramp-up Trip} automaton detects blocks with statuses 3, 11, or 512. These statuses adhere to CAEN standards, where bits indicate conditions: 1 for channel ON, 2 for ramp-up, 8 for over-current (OVC), and 512 for trip. By categorizing trips into operational and ramp-up events, these automata store the information in the RPC condition schema for subsequent correlation studies and historical event analysis.

Another automaton monitors over-current spikes, capturing instances where the hardware status OVC bit is triggered. These occurrences are logged in the same evolution table of the database for use in correlation analyses and real-time notifications, enhancing the robustness of RPC monitoring and diagnostics.
\subsection{Active Channels}
There are 123,432 RPC readout channels in the Run-II and Run-III operation system. Each electronic channel reads the induced signal on a copper strip from the readout plane situated between two gas gap layers \cite{cms_rpc_pugliese}. Channels may be masked due to noise, hardware issues, or malfunctions. The individual strip \textit{enable bit}, stored in the configuration database, is loaded onto the hardware during CMS detector configuration.

To monitor active channels, a dedicated automaton, \textit{Active Channels}, was introduced into the RPC automation framework. This automaton records the number of active strips and timestamps any changes for every individual strip. Tracking this parameter is invaluable for assessing the RPC system's engagement and performance during operation.
\subsection{Machine Learning Automata}
A series of machine learning-based automata analyze preprocessed data on currents and environmental and operational conditions from the RPC automation framework. These automata model system behavior to enable accurate predictions of future performance. In line with the project’s full-circle automation ideology, detailed in Section \ref{idea}, discrepancies between real-time and predicted currents trigger notifications for prompt operator intervention. Eventually, all manual fine-tuning will be replaced by an automated system, which will adjust detector parameters using predefined \textit{DCS} data points.
\section{Conclusion}
This paper presents a novel framework for real-time processing and analysis of raw condition data from the non-physics event bus of the CERN Compact Muon Solenoid experiment. Leveraging advanced data streaming techniques, the framework investigates correlations between key Resistive Plate Chamber operational parameters — such as currents and rates— with the LHC instantaneous luminosity and environmental factors. It also tracks the evolution of these parameters over time and predicts future detector behavior by modeling performance based on identified correlations.
\section*{Acknowledgments}
We gratefully acknowledge Osvaldo M. Colin for his invaluable contributions in designing and implementing the RPC automation framework in Java, as well as for developing a significant portion of the associated methods and automata that form the foundation of this work.

We congratulate our colleagues in the CERN accelerator departments for the excellent performance of the LHC and thank the technical and administrative staffs at CERN and at other CMS institutes for their contributions to the success of the CMS effort. In addition, we gratefully acknowledge the computing centers and personnel of the Worldwide LHC Computing Grid and other centers for delivering so effectively the computing infrastructure essential to our analyses. Finally, we acknowledge the enduring support for the construction and operation of the LHC, the CMS detector, and the supporting computing infrastructure provided by the following funding agencies: FWO (Belgium); CNPq, CAPES and FAPERJ (Brazil); MES and BNSF (Bulgaria); CERN; CAS, MoST, and NSFC (China); MINCIENCIAS (Colombia); CEA and CNRS/IN2P3 (France); SRNSFG (Georgia); IPM (Iran); INFN (Italy); MSIP and NRF (Republic of Korea); BUAP, CINVESTAV, CONACYT, LNS, SEP, and UASLP-FAI (Mexico); PAEC (Pakistan); DOE and NSF (USA). 

\bibliographystyle{apalike}

\end{document}